\documentclass[showpacs,preprintnumbers,amsmath,amssymb,twocolumn,nobibnotes,longbibliography,nofootinbib]{revtex4-2}
\usepackage[english]{babel}
\usepackage{mathtools}
\usepackage{graphicx}
\usepackage{mathrsfs}
\usepackage{amssymb}
\usepackage{xcolor}
\usepackage[colorlinks=true,citecolor=blue,urlcolor=blue]{hyperref}
\usepackage{amsmath}
\usepackage{bm}
\usepackage{physics}
\newcommand{\mysection}[1]{\textcolor{blue}{\textit{#1}.}}

\begin{document}

\title{Edge spin galvanic effect in  altermagnets}

\author{L. E. Golub}
\affiliation{Institute of Theoretical Physics and Halle-Berlin-Regensburg Cluster of Excellence CCE, University of Regensburg, 93040 Regensburg, Germany}

\begin{abstract}
The edge spin galvanic effect (ESGE) in $d$-wave altermagnets is proposed. 
ESGE is a creation of an electrical current flowing along the edge of the sample, which is driven by the spin orientation of charge carriers.
The ESGE current is formed owing to the altermagnetic spin splitting and the scattering of carriers by the edge of the sample.  The current is sensitive to the orientation of the edge in respect to the main axes of the altermagnet. The edge spin galvanic current reverses its direction upon a reversal of the non-equilibrium spin direction or the N\'eel vector.
We also propose the pure spin edge photocurrent excited by polarized radiation and  formed at the edges of a sample. 
Its dependence on the radiation polarization and frequency is analyzed.
The application of an external magnetic field converts this pure spin photocurrent into an electric current along the edge.
\end{abstract}

\maketitle

\mysection{Introduction}
The spin galvanic effect is a conversion of a nonequilibrium electron spin polarization into electric current~\cite{Ganichev2002}. This remarkable phenomenon which lies at the heart of spintronics 
has been observed in many semiconductor and metal systems~\cite{Ivchenko2005,Ivchenko2017,Ganichev2024}.
For a possibility of this spin to the current conversion, the system symmetry should be low enough: The spin-galvanic effect is possible only in gyrotropic systems.
This ensures the odd in momentum spin splitting of the bandstructure which is a prerequisite of the spin galvanic effect. 

Altermagnets---a rapidly developing class of condensed-matter systems---are known by their large nonrelativistic spin splittings which exceed the spin-orbit splittings in traditional systems by a few orders in magnitude~\cite{Smejkal2022,Smejkal2022a,Jungwirth2025}. However, despite of the spin splitting values up to 1eV, the altermagnets are centrosymmetric media, where the spin galvanic effect is forbidden by symmetry. Therefore,  interconversion of the electric current and spin is only possible in altermagnets in the nonlinear regime, where the current induced spin orientation is proportional to the second or higher even powers of the current, depending on the type of altermagnet~\cite{Golub2025}.

Despite these symmetry constraints, we show below that the spin galvanic currents can flow along edges of $d$-wave altermagnetic samples. We name this phenomenon the edge spin galvanic effect (ESGE).

\begin{figure}[t]
	\centering 
		\includegraphics[width=\linewidth]{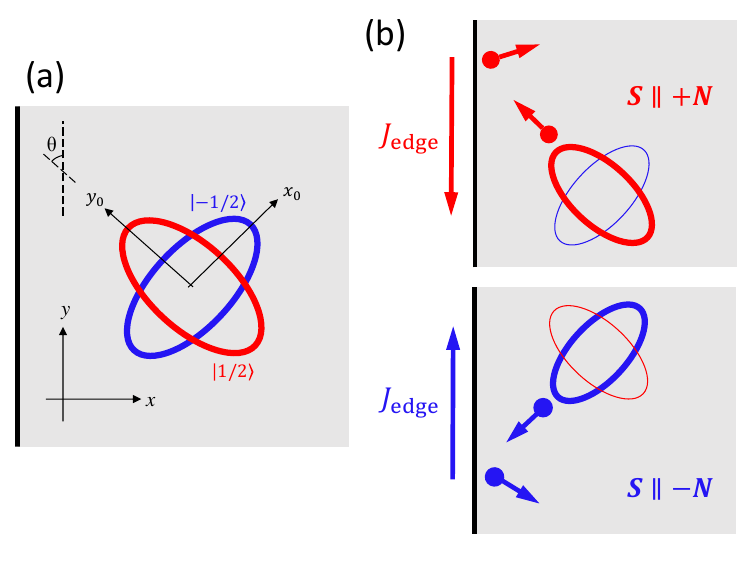}
	\caption{(a): Semi-infinite $d$-wave altermagnet with the edge along $y$ axis. Red and blue curves show the Fermi contours for two spin states with the main axes $(x_0,y_0)$ rotated by angle $\theta$ in respect to $(x,y)$.
	(b): Edge spin galvanic current formation. 
In the presence of a nonequilibrium spin $\bm S$ parallel to the N\'eel vector $\bm N$, electrons from each spin subband flow to the sample edge at some preferred angle. 
The asymmetry in the electron momentum distribution emerges due to scattering off the edge, thus forming the edge current.
At opposite spin orientation the edge current reverses its direction.
	}
	\label{Fig_setup_edge_SGE}
\end{figure}

\mysection{ESGE current}
We consider a spin-polarized semi-infinite plane $d$-wave altermagnet with the edge along $y$ direction, Fig.~\ref{Fig_setup_edge_SGE}(a). Spin polarization means a preferable occupation of one of spin subbands. 
We show that an electric current flows along the edge in this setup, Fig.~\ref{Fig_setup_edge_SGE}(b).
This effect is described by the phenomenological relation
\begin{equation}
\label{J_edge_phenom}
J_{\rm edge} = \Xi {S}_N,
\end{equation}
where $J_{\rm edge}$ is the edge electric current, 
${S}_N$ is the nonequilibrium electron spin component along the N\'{e}el vector $\bm N$, 
and the coefficient $\Xi$ reflects the altermagnetic order and 
changes its sign upon a reversal of the N\'{e}el vector.

Qualitatively, the two microscopic ingredients result to the ESGE current. First, the spin-oriented carriers in a $d$-wave altermagnet have {\em momentum alignment}, which means that their distribution is anisotropic in the momentum space and varies with the momentum direction as a second angular harmonics. 
This kind of momentum distribution can be created via absorption of electromagnetic radiation in any conducting system, including altermagnets~\cite{Golub2025}. However, in $d$-wave altermagnets symmetry directly relates the momentum alignment with spin. 
The second ground for the ESGE is scattering of charge carriers by the edge of the sample, Fig.~\ref{Fig_setup_edge_SGE}(b). The presence of the edge removes the spatial inversion (the corresponding vector is the normal to the edge) and makes the spin galvanic current symmetry-allowed. We show that a combination of momentum alignment and edge scattering in $d$-wave altermagnets results in the ESGE current.

The energy spectra of electrons 
in two spin subbands
in a $d$-wave altermagnet have the form
\begin{equation}
\label{energy_disp}
\varepsilon_{\bm k}^\pm = \varepsilon_k \pm \beta  (k_{x_0}^2 - k_{y_0}^2),
\quad \varepsilon_k={\hbar^2 k^2\over 2m} .
\end{equation}
Here $\bm k$ is the two-dimensional wavevector, $m$ is the effective mass, 
$x_0,y_0$ are the main axes of the $d$-wave altermagnet, Fig.~\ref{Fig_setup_edge_SGE}(a), and $\beta$ describes the altermagnetic order.

We calculate the edge current~\eqref{J_edge_phenom} in the linear order in $\beta$ assuming $\beta \ll \hbar^2/m$.
In the presence of spin pumping, the generation rates in the spin subbands are given by $G^\pm_{\bm k} = 
\pm 2\dot{S}_N f_0(\varepsilon_{\bm k}^\pm)/ n$, where  $f_0(\varepsilon)$ is the Fermi-Dirac distribution, $n$ is the electron concentration, and $\dot{\bm S}$
is the spin generation rate~\cite{Ivchenko2005}.
In the steady-state conditions, $\dot{S}_N=S_N/\tau_s$, where $\tau_s$ is the spin relaxation time for the corresponding spin component.
The generation rates
contain the spin-independent part contributing to the current:
\begin{equation}
\label{generation}
G_{\bm k} =  {2S_N \over n\tau_s}  \beta  (k_{x_0}^2 - k_{y_0}^2) 
\dv{f_0(\varepsilon_k)}{\varepsilon_k}.
\end{equation}
Equation~\eqref{generation}
describes the alignment of electron momenta under spin orientation that was discussed above.
Microscopically, it occurs during the process of spin relaxation.

We use the kinetic theory and
introduce the electron distribution function
$f(\bm k, x)$ depending on the wavevector and the coordinate $x$ perpendicular to the edge.
The correction to the distribution function satisfies the Boltzmann kinetic equation. At steady state spin pumping, it has the following form
\begin{equation}
\label{kin_eq}
v_x \pdv{\delta f}{x} = G_{\bm k} 
- {\delta f \over \tau}.
\end{equation}
Here $\delta f= f(\bm k, x)-f_0(\varepsilon_k)$ is the nonequilibrium correction to the distribution function, $\bm v = \hbar \bm k/m$ is the electron velocity,
and $\tau$ is the relaxation time.
Since the spin splitting is accounted for in the generation rate $G_{\bm k}$,  we should disregard the $\beta$-dependent corrections to the velocities and collision integral.

The electric current along the edge is calculated by integrating  the current density $j_y(x)$
\begin{equation}
\label{J_edge_j}
J_{\rm edge}= \int_0^\infty \dd x j_y(x), \quad  j_y(x)=2e\sum_{\bm k}v_y \delta f(\bm k, x),
\end{equation}
where the factor 2 accounts for the
spin degeneracy.
Solving the kinetic Eq.~\eqref{kin_eq}, see Supplemental Material for details, we obtain the edge spin-galvanic current in the form of Eq.~\eqref{J_edge_phenom}
\begin{equation}
\label{J_edge_result}
J_{\rm edge} = \Xi {S}_N, \quad \Xi=  \beta k_{\rm F}^2 {e\tau^2 \over m \tau_s}\sin{2\theta} .
\end{equation}
Here $k_{\rm F}$ is the Fermi wavevector, and specular scattering from the edge is assumed.
This expression shows that 
the edge current flows in $d$-wave altermagnetic samples at spin pumping provided the main axes are not parallel to the edges.
The value $\Xi$ 
is even under time reversal because  the altermagnetic order parameter $\beta$ is odd.

For $\beta k_{\rm F}^2=1$~eV, $\tau=1$~ps, $m=m_0$, ${S}_N = 10^{12}$~cm$^{-2}$ and $\tau_s=100$~ps
we obtain an estimate $J_{\rm edge} \approx 1$~$\mu$A. This is the order of the edge currents that was measured in the optical experiments~\cite{Glazov2014,Candussio2020,Candussio2021}.

\begin{figure}[t]
	\centering 
		\includegraphics[width=\linewidth]{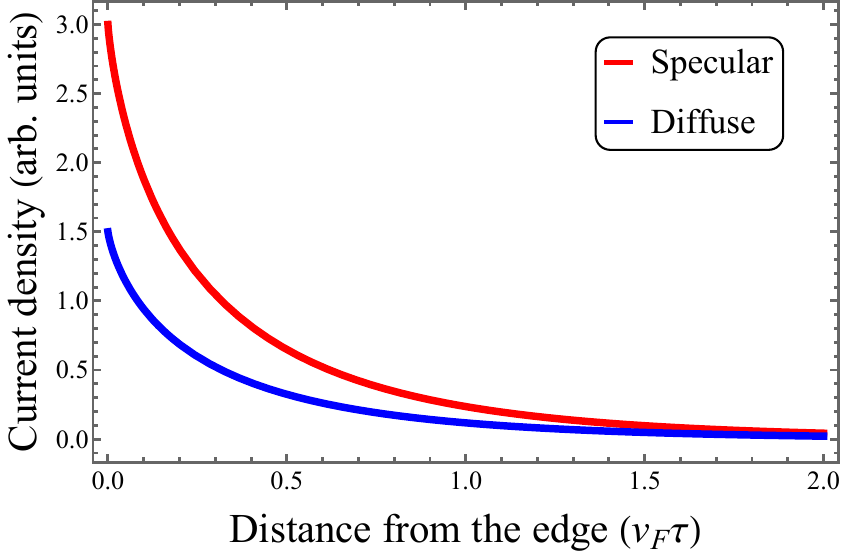}
	\caption{Spatial distribution of the ESGE current density near the sample edge for specular and diffuse edge scattering. The current density is normalized by $J_{\rm edge}/(v_{\rm F}\tau)$, where $J_{\rm edge}$ is the total current at specular scattering Eq.~\eqref{J_edge_result}.
	}
	\label{Fig_spin_current_vs_x}
\end{figure}

The edge current~\eqref{J_edge_result} is formed in the vicinity of the edge and exists within a stripe whose width is of the order of the electron mean free path. 
The ESGE current density $j_y(x)$, Eq.~\eqref{J_edge_j},
is calculated in Supplemental Material.
The spatial distribution $j_y(x)$ is presented in Fig.~\ref{Fig_spin_current_vs_x} for two types of edge scattering.
The ESGE current density near the sample edge  drops rapidly with $x$ and almost vanishes at $x>v_{\rm F}\tau$, where $v_{\rm F}$ is the Fermi velocity.

The ESGE current~\eqref{J_edge_result} is derived for steady-state spin generation. If the electron spin
varies in time and the characteristic time of variation is long as
compared to the momentum and energy relaxation times
then $\dot{S}_N$ is substituted by $\dot{S}_N  - {\dd S_N(t)/ \dd t}$~\cite{Ivchenko2017,Ganichev2024}.
Particularly, under a short-pulse spin excitation at $t = 0$,  the edge spin-galvanic current at $t > 0$ has the form
\begin{equation}
\label{J_t}
J_{\rm edge}(t) = -\Xi \tau_s {\dd S_N(t)\over \dd t}.
\end{equation}
In the presence of magnetic field perpendicular to the 
N\'eel vector,
the spin dynamics is described by the damped oscillatory behavior
\begin{equation}
\label{s_t}
S_N(t)=S_N(0) {\rm e}^{-t/\tau_s} \cos(\Omega_L t),
\end{equation}
where $\Omega_L$ is the Larmor precession frequency.
For the magnetic field oriented in the $(x_0,y_0)$ plane 
the Lorentz force does not appear,
and the ESGE current time dependence is determined solely by the spin dynamics. From Eqs.~\eqref{J_t} and~\eqref{s_t} we obtain 
\begin{equation}
\label{J_SGE_time-dep}
J_{\rm edge}(t) = J_{\rm edge}(0) {\rm e}^{-t/\tau_s} \qty[ \cos(\Omega_L t)+\Omega_L\tau_s\sin(\Omega_L t)].
\end{equation}
This regime of ESGE can also 
be understood in terms of 
the coherent trembling motion (Zitterbewegung) of spin-polarized electrons in the external magnetic field~\cite{Ivchenko2017,Tarasenko2018}.

\begin{figure}[t]
	\centering 
		\includegraphics[width=0.8\linewidth]{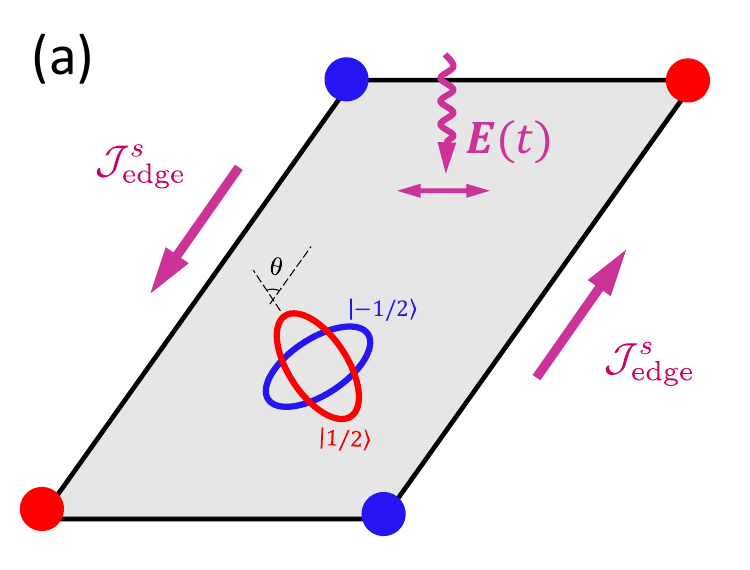}
				\includegraphics[width=0.8\linewidth]{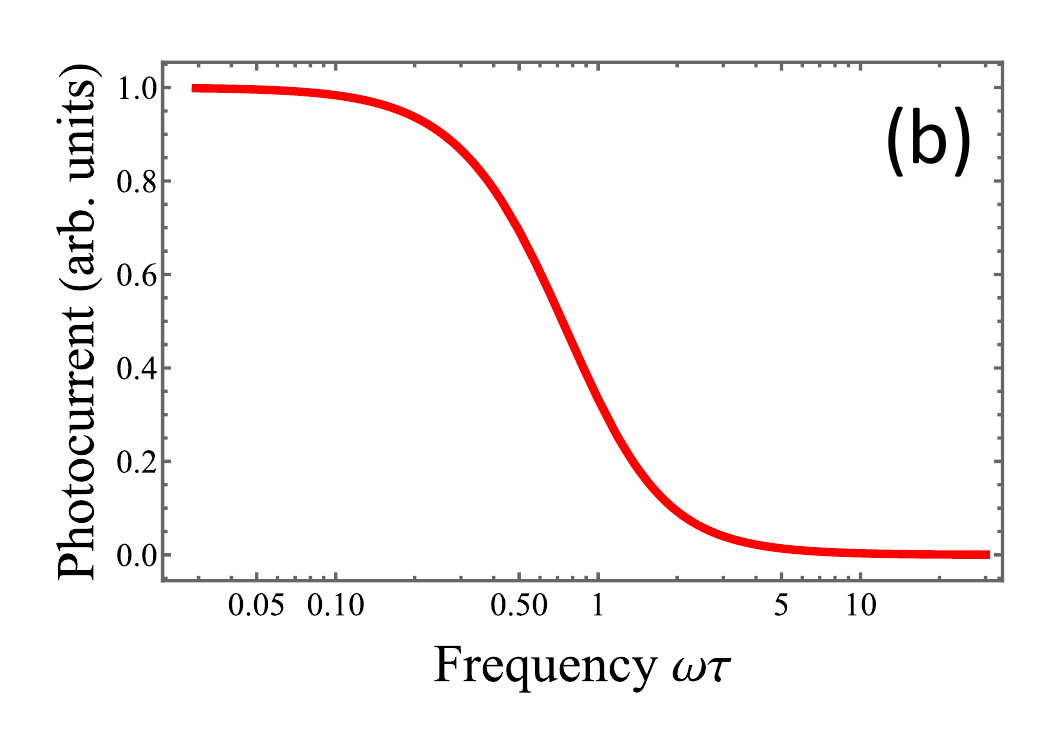}
	\caption{Pure spin edge photocurrent 	in a $d$-wave altermagnet. (a) The current is maximal at the  radiation polarization perpendicular to the edge.
It varies as $\propto \sin{2\theta}$ with the angle $\theta$ between the edge and the main axes of the $d$-wave altermagnet, Eq.~\eqref{J_s}. The spin-up and spin-down carriers shown by red and blue circles are accumulated at the corners of the sample. (b) Pure spin edge photocurrent dependence on the radiation frequency.
	}
	\label{Fig_spin_current}
\end{figure}

\mysection{Pure spin edge photocurrent}
Let us consider an effect of electromagnetic radiation on altermagnets.
Absorption of linearly-polarized or unpolarized radiation may result in the dc edge photocurrent in any material~\cite{Glazov2014,Candussio2020,Durnev2020,Durnev2021,Candussio2021}. 
However, the edge current is absent when the radiation is polarized perpendicular to the edge.
Below we demonstrate that 
this radiation results in the flow of electrons with opposite spins in the opposite directions along the edge. Thus, 
light absorption in $d$-wave altermagnets generates
the dc pure spin edge current 
\[
\mathcal J_{\rm edge}^s=(J_{\rm edge}^+ - J_{\rm edge}^-)/2.
\]
This is illustrated in Fig.~\ref{Fig_spin_current}(a).
Here the charge currents in the subbands are given by
\begin{equation}
J_{\rm edge}^\pm=e\sum_{\bm k} \int_0^\infty \dd x v_y^\pm(\bm k) f_\pm(\bm k, x),
\end{equation}
where $f_\pm(\bm k, x)$ are the dc electron distribution functions in the spin subbands
quadratic in the amplitude of the radiation' electric field,
and $\bm v^\pm(\bm k)=\bm \nabla_{\bm k}\varepsilon_{\bm k}^\pm/\hbar$ are the velocities.

The functions $f_\pm(\bm k, x)$ satisfy the kinetic equations with complex collision integrals 
due to the presence of the spin splitting in the elastic scattering probabilities~\cite{Ivchenko2005,Golub2011}.
However, changing the variables in each subband $\bm k \to \bm q$ where $q_{x_0}=k_{x_0}(1\pm \beta m/\hbar^2)$, $q_{y_0}=k_{y_0}(1\mp \beta m/\hbar^2)$, we obtain the isotropic spectra $\varepsilon_q=(\hbar q)^2/ (2m)$ in both subbands. This simplifies greatly the collision integral because the elastic scattering in the $\bm q$-space is reduced to averaging over directions of $\bm q$. As a result, 
the kinetic equation for $f_\pm(\bm q, x)=f_0(\varepsilon_q)+\delta f_\pm(\bm q, x)$ 
in the presence of the linearly-polarized radiation with the electric field $\bm E(t) = \bm E [\exp(-i\omega t)+\exp(i\omega t)]$
has the following form
\begin{equation}
\label{kin_eq_E_field1}
\pdv{\delta f_\pm}{t} + v_x^\pm(\bm q) \pdv{\delta f_\pm}{x} + {e\over \hbar} {\bm E}^\pm(t)\cdot \bm \nabla_{\bm q}f_\pm= - {\delta f_\pm \over \tau}.
\end{equation}
Here $v_x^\pm(\bm q)$ is a component of the velocity $\bm v^\pm(\bm k)$ expressed via $\bm q$: $\bm v^\pm(\bm q) = \hbar \bm q/m \pm (\beta /\hbar)\hat{\mathcal M} \bm q$, and we have rewritten $\bm E(t)\cdot \bm \nabla_{\bm k}$ as ${\bm E}^\pm(t)\cdot \bm \nabla_{\bm q}$, where ${\bm E}^\pm = \bm E \pm (m\beta /\hbar^2)\hat{\mathcal M} \bm E$
and $\hat{\mathcal M} 
= \begin{pmatrix}
\cos{2\theta}&\sin{2\theta}\\
\sin{2\theta}&-\cos{2\theta}
\end{pmatrix}.
$

The correction to the distribution function in each subband, $\delta f_\pm$, is found by solving Eq.~\eqref{kin_eq_E_field1} as a Taylor expansion
\begin{equation}
\delta f_\pm(\bm k, t, x) =f_\pm^{(1)} {\rm e}^{-i\omega t} + {f_\pm^{(1)}}^*  {\rm e}^{i\omega t} +f_\pm^{(2)},
\end{equation}
where $f_\pm^{(1)} \propto E$, and $f_\pm^{(2)} \propto {E}^2$.
We find
that the edge currents are opposite in two spin subbands,
${J_{\rm edge}^\pm = \pm \mathcal J_{\rm edge}^s}$, see Supplemental Material for details of calculations.
The  pure spin photocurrent  flows along the edges in $d$-wave altermagnets  for the polarization perpendicular to the edge, $\bm E \parallel x$:
\begin{equation}
\label{J_s}
\mathcal J_{\rm edge}^s =-2\beta  \sin{2\theta}{n(e\tau)^3 [3+(\omega \tau)^2]\over m\hbar^2[1+(\omega \tau)^2]^2} E_x^2.
\end{equation}
The frequency dependence of the edge pure spin photocurrent is shown in Fig.~\ref{Fig_spin_current}(b).
It shows that the spin current is almost frequency-independent at low frequencies but drops rapidly at $\omega > 1/\tau$.

Application of the magnetic field $\bm B \parallel \bm N$ converts the edge spin current $\mathcal J_{\rm edge}^s$ into the electric edge current. If the N\'eel vector $\bm N$ lies in the $(x_0,y_0)$ plane, 
then the Lorentz force is absent,
and 
the electric photocurrent appears exclusively due to the spin current conversion~\cite{Ganichev2006}. The edge photocurrent $J_{\rm edge} =4(S_{N}/n)\mathcal J_{\rm edge}^s$ reads
\begin{equation}
J_{\rm edge} =2B \beta  \sin{2\theta}{g\mu_{\rm B} (e\tau)^3 [3+(\omega \tau)^2]\over \pi \hbar^4[1+(\omega \tau)^2]^2} E_x^2.
\end{equation}
Here $g$ is the Land\'e factor, and we used the expressions for the equilibrium spin density $S_{N}=-(g\mu_{\rm B}B/4\varepsilon_{\rm F})n$ and concentration $n=m\varepsilon_{\rm F}/(\pi \hbar^2)$ in degenerate two-dimensional systems with Fermi energy $\varepsilon_{\rm F}$.
This magnetoinduced photocurrent at $\bm E \parallel x$ is different from that in two-dimensional non-magnetic systems, where it is caused by the Lorentz force in the out of plane field~\cite{Candussio2020,Durnev2020}. Here, the electric current is formed due 
conversion of the pure spin photocurrent  excited in the absence of magnetic field.

\mysection{Concluding remarks} We developed a theory of the edge spin galvanic effect in two-dimensional $d$-wave altermagnets. We also studied the pure spin edge photocurrent which is excited by a polarized light and can be converted into the edge electric current by application of the magnetic field. 
Analogous phenomena exist in three-dimensional $d$-wave altermagnets: the surface spin galvanic current and the surface pure spin photocurrent. They are caused by the scattering of electrons off the surface in bulk samples.

We note that the edge currents are regularly measured in experiments, e.g. under excitation 
by terahertz radiation~\cite{Glazov2014,Candussio2020,Candussio2021}. The spin pumping of altermagnets is also intensively studied~\cite{Sun2023,Guo2024}. A combination of these techniques allows detecting the proposed ESGE current as well as the edge spin photocurrent.

\mysection{Acknowledgments} Author thanks E.~L.~Ivchenko and S.~A.~Tarasenko for discussions. 
This work was funded by the German Research Foundation (DFG) as part of the German Excellence Strategy -- EXC3112/1 -- 533767171 (Center for Chiral Electronics).

\bibliography{Library_ESGE}

\onecolumngrid

\newpage

\begin{center}
\makeatletter
{\large\bf{Supplemental Material for\\``\@title''}}
\makeatother
\end{center}

\let\oldsec\section

\renewcommand{\thesection}{S\arabic{section}}
\renewcommand{\section}[1]{\oldsec{#1}}
\renewcommand{\thepage}{S\arabic{page}}
\renewcommand{\theequation}{S\arabic{equation}}
\renewcommand{\thefigure}{S\arabic{figure}}

\setcounter{page}{1}
\setcounter{section}{0}
\setcounter{equation}{0}
\setcounter{figure}{0}

\section{Edge spin galvanic current calculation}

Multiplying Eq.~\eqref{kin_eq} by $e v_y$, summing over 
$\bm k$ and spin subbands, and integrating over $x$ we obtain
\begin{equation}
\label{J_edge_eq}
2e\sum_{\bm k} v_x v_y \qty[\delta f(\bm k, \infty)-\delta f(\bm k, 0)] = -{J_{\rm edge} \over \tau}.
\end{equation}
We took into account that the spin pumping is not accompanied by electric current in the bulk: the generation rate~\eqref{generation} satisfies 
$\sum_{\bm k}v_y G_{\bm k}=0$.
For the specular edge scattering, the term with  $\delta f(\bm k, 0)$ cancels because it is even in $k_x$ while $v_x$, calculated at $\beta=0$, is odd in $k_x$. Therefore the edge current is determined by the distribution function far from the edge, $\delta f(\bm k, \infty)$. It is found from Eq.~\eqref{kin_eq} with zero left-hand side:  
$\delta f(\bm k, \infty)= G_{\bm k}\tau $.
This gives according to Eq.~\eqref{generation}
\begin{equation}
\delta f(\bm k, \infty) = 2{\dot{S}_N \over n} \tau \beta  (k_{x_0}^2 - k_{y_0}^2) f_0'(\varepsilon_k),
\end{equation}
where $f_0'(\varepsilon_k)=\dd f_0(\varepsilon_k)/\dd \varepsilon_k$, and
\begin{equation}
\label{J_edge_1}
J_{\rm edge} = -4\beta {\dot{S}_N \over n} e\tau^2\sum_{\bm k}v_x v_y   (k_{x_0}^2 - k_{y_0}^2) f_0'(\varepsilon_k).
\end{equation}
Then we pass to the coordinates 
$(x_0,y_0)$:
\begin{equation}
\label{xy_x0y0_coordinates}
v_x v_y = v_{x_0} v_{y_0}\cos{2\theta} + {v_{x_0}^2-v_{y_0}^2\over 2}\sin{2\theta}.
\end{equation}
Since Eq.~\eqref{J_edge_1}
contains $k_{x_0}^2 - k_{y_0}^2$, only the second term here contributes to $J_{\rm edge}$:
\begin{multline}
J_{\rm edge} = -4\beta {\dot{S}_N \over n} e\tau^2\sin{2\theta} \sum_{\bm k}{v_{x_0}^2-v_{y_0}^2\over 2} (k_{x_0}^2 - k_{y_0}^2) f_0'(\varepsilon_k)
 = -2\beta {\dot{S}_N \over n} e\tau^2\sin{2\theta} \sum_{\bm k}k^2v^2\cos^2{2\varphi_{\bm k}} f_0'(\varepsilon_k)
\\ = -\beta {\dot{S}_N \over n} e\tau^2\sin{2\theta} \sum_{\bm k}k^2v^2 f_0'(\varepsilon_k)
= \beta k_{\rm F}^2 \dot{S}_N e\tau^2\sin{2\theta} {v_{\rm F}^2 \over 2\varepsilon_{\rm F}}
 = \dot{S}_N \beta k_{\rm F}^2 {e\tau^2 \over m}\sin{2\theta} .
\end{multline}
We assumed here a degenerate statistics, 
and $\varepsilon_{\rm F}$, $k_{\rm F}$  and $v_{\rm F}$
are the Fermi energy, wavevector and velocity.


\section{ESGE current density}

The density of ESGE current is calculated by Eq.~\eqref{J_edge_j}:
\begin{equation}
j_y(x) = 2e\sum_{\bm k} v_y \delta f(\bm k, x).
\end{equation}
The asymmetric in respect to the change $k_y \to -k_y$ part of $\delta f(\bm k, x)$ is given by~\cite{Durnev2021}
\begin{equation}
\delta f(\bm k, x) = G_{\bm k}\tau+\Theta(k_x)\qty(\zeta G_{-k_x,k_y}-G_{k_x,k_y})\tau\exp(-{x\over v_x\tau}),
\end{equation}
where $\zeta=1$ and $\zeta=0$ for specular and diffuse edge scattering, respectively.
It follows from Eq.~\eqref{generation} and the relation
\begin{equation}
k_{x_0}^2-k_{y_0}^2 = 2k_xk_y \sin{2\theta} + (k_x^2-k_y^2)\cos{2\theta},
\end{equation}
that the asymmetric in $k_y$ part of the generation rate reads
\begin{equation}
G_{\bm k} = 4\beta {\dot{S}_N \over n} \sin{2\theta} k_xk_y f_0'(\varepsilon_k).
\end{equation}
This yields
\begin{multline}
j_y(x) 
= -8\beta {\dot{S}_N \over n} \sin{2\theta} e\tau(1+\zeta)\sum_{\bm k} v_y k_xk_y f_0'(\varepsilon_k)\Theta(k_x)\exp(-{x\over v_x\tau})
\\ ={J_{\rm edge} \over v_{\rm F}\tau} {8\over \pi}\int_{-\pi/2}^{\pi/2} \dd \varphi \sin^2\varphi \cos\varphi \exp(-{x\over v_{\rm F}\tau\cos\varphi})
= {J_{\rm edge}  \over v_{\rm F}\tau} {16\over \pi}\int_1^{\infty} \dd z {(z^2-1)^{3/2}\over z^5}\exp(-z {x\over v_{\rm F}\tau}).
\end{multline}
Here we took into account that  the total electric current $J_{\rm edge}$ for specular edge scattering is twice larger than for diffuse. As a result, 
the ratios $j_y(x)/J_{\rm edge}$ are equal, and
this expression is valid for both types of scattering.
Numerical integration of the above expression gives the $x$-dependence of the ESGE current density shown in Fig.~\ref{Fig_spin_current_vs_x}.

\section{Pure spin edge photocurrent calculation}

We consider the polarizations $\bm E \parallel x$ or $y$, where the electric photocurrent is absent. 
We solve Eq.~\eqref{kin_eq_E_field1}:
\begin{equation}
\pdv{\delta f_\pm}{t} + v_x^\pm(\bm q) \pdv{\delta f_\pm}{x} + {e\over \hbar} {\bm E}^\pm(t)\cdot \bm \nabla_{\bm q}f_\pm= - {\delta f_\pm \over \tau}
\end{equation}
with $\bm v^\pm(\bm q) = \hbar \bm q/m \pm (\beta /\hbar)\hat{\mathcal M} \bm q$, ${\bm E}^\pm = \bm E \pm (m\beta /\hbar^2)\hat{\mathcal M} \bm E$, and
$\hat{\mathcal M} 
= \begin{pmatrix}
\cos{2\theta}&\sin{2\theta}\\
\sin{2\theta}&-\cos{2\theta}
\end{pmatrix}.
$

Multiplying Eq.~\eqref{kin_eq_E_field1} by $e v_y^\pm(\bm q)$, summing over $\bm q$, and integrating over $x$ we obtain for the steady state photocurrents in the spin subbands
\begin{equation}
\label{J_1}
J_{\rm edge}^\pm = -e\tau\sum_{\bm q}v_x^\pm(\bm q) v_y^\pm(\bm q) \qty[f_\pm^{(2)} (\bm q,\infty)-f_\pm^{(2)} (\bm q,0)]
 - {e^2\tau\over \hbar}\sum_{\bm q}v_y^\pm(\bm q) \qty(\bm E^\pm\cdot \bm \nabla_{\bm q}) \int_0^\infty \dd x \qty[f_\pm^{(1)} (\bm q,x)+c.c.].
\end{equation}
To calculate the last term in Eq.~\eqref{J_1}, we derive from Eq.~\eqref{kin_eq_E_field1} the equation for $f_\pm^{(1)} (\bm k,x)$:
\begin{equation}
\label{eq_f_1}
-i\omega f_\pm^{(1)}+ v_x^\pm \pdv{f_\pm^{(1)}}{x}+ {e\over \hbar} \qty(\bm E^\pm\cdot \bm \nabla_{\bm q})f_0(\varepsilon_{\bm q})= - {f_\pm^{(1)} \over \tau}.
\end{equation}
Summation of this equation over $\bm q$ gives
\begin{equation}
\sum_{\bm q} f_\pm^{(1)} = -\tau_\omega \sum_{\bm q} v_x^\pm \pdv{f_\pm^{(1)}}{x}, \quad \tau_\omega = {\tau\over 1-i\omega\tau},
\end{equation}
and, hence,
\begin{equation}
\sum_{\bm q}\int_0^\infty \dd x f_\pm^{(1)} (\bm q,x) = -\tau_\omega \sum_{\bm q} v_x^\pm \qty[f_\pm^{(1)} (\bm q,\infty)-f_\pm^{(1)} (\bm q,0)].
\end{equation}
Then we obtain from Eq.~\eqref{J_1} the edge current in each subband
\begin{equation}
\label{J_2}
J_{\rm edge}^\pm =  -e\tau\sum_{\bm q}v_x^\pm v_y^\pm \qty[f_\pm^{(2)} (\bm q,\infty)-f_\pm^{(2)} (\bm q,0)]
 - {e^2\tau\over \hbar}\qty(\bm E^\pm\cdot \bm \nabla_{\bm q})v_y^\pm 
\qty{ \tau_\omega \sum_{\bm q} v_x^\pm \qty[f_\pm^{(1)} (\bm q,\infty)-f_\pm^{(1)} (\bm q,0)] +c.c.}.
\end{equation}
Here we took into account that $\bm v^{\pm}$ is linear in $\bm q$, and, hence, $\partial_{q_i} v_j^\pm$ are constants.

The corrections far from the edge, $f_\pm^{(1,2)} (\bm q,\infty)$, are found from a homogeneous version of Eq.~\eqref{kin_eq_E_field1}:
\begin{equation}
\label{kin_eq_E_field_infty}
\pdv{\delta f_\pm(\bm q, \infty)}{t} + {e\over \hbar} \bm E^{\pm}(t)\cdot \bm \nabla_{\bm q}f_\pm(\bm q, \infty)= - {\delta f_\pm(\bm q, \infty) \over \tau}.
\end{equation}
Substitution of the equilibrium distribution into the field term of Eq.~\eqref{kin_eq_E_field_infty} gives
\begin{equation}
\label{f_1_infty}
f_\pm^{(1)} (\bm q,\infty) = -{e\over \hbar}\tau_\omega \qty(\bm E^{\pm}\cdot \bm \nabla_{\bm q}) f_0(\varepsilon_q).
\end{equation}
The next iteration of the kinetic Eq.~\eqref{kin_eq_E_field_infty} yields the second order correction in the form
\begin{equation}
f_\pm^{(2)} (\bm q,\infty) = {2(e\tau)^2\over \hbar^2[1+(\omega \tau)^2]} \qty(\bm E^\pm\cdot \bm \nabla_{\bm q})^2f_0(\varepsilon_q).
\end{equation}
Now we can calculate the sums in Eq.~\eqref{J_2} containing $f_\pm^{(1,2)} (\bm q,\infty)$. 
We start with $f_\pm^{(2)} (\bm q,\infty)$:
\begin{equation}
\label{J_f_2_infty}
-e\tau\sum_{\bm q}v_x^\pm v_y^\pm  f_\pm^{(2)} (\bm q,\infty)
=- {2(e\tau)^3\over \hbar^2[1+(\omega \tau)^2]} \sum_{\bm q} v_x^\pm v_y^\pm \qty(\bm E^\pm\cdot \bm \nabla_{\bm q})^2f_0(\varepsilon_q)
=- {2(e\tau)^3\over \hbar^2[1+(\omega \tau)^2]} \sum_{\bm q}f_0(\varepsilon_q) \qty(\bm E^\pm\cdot \bm \nabla_{\bm q})^2 v_x^\pm v_y^\pm .
\end{equation}
The velocities are given by
\begin{equation}
\label{vx}
v_x^\pm = {\hbar q_x \over m} \qty(1\pm {m\beta\over \hbar^2}\cos{2\theta}) \pm {\beta\over \hbar}\sin{2\theta}q_y,
\quad
v_y^\pm = {\hbar q_y \over m} \qty(1 \mp {m\beta\over \hbar^2}\cos{2\theta}) \pm {\beta\over \hbar}\sin{2\theta}q_x.
\end{equation}
Then (for $E_xE_y=0$)
\begin{equation}
 v_x^\pm v_y^\pm \approx \qty({\hbar \over m})^2 q_xq_y \pm {\beta \over m}\sin{2\theta}(q_x^2+q_y^2),
\quad \qty(\bm E^\pm\cdot \bm \nabla_{\bm q})^2
v_x^\pm v_y^\pm = \pm{4\beta \over m}\sin{2\theta} (E_x^2+E_y^2),
\end{equation}
and we get from Eq.~\eqref{J_f_2_infty}:
\begin{equation}
-e\tau\sum_{\bm q}v_x^\pm v_y^\pm  f_\pm^{(2)} (\bm q,\infty) 
=- {2(e\tau)^3\over \hbar^2[1+(\omega \tau)^2]} \sum_{\bm q}f_0(\varepsilon_q) \qty[\pm{4\beta \over m}\sin{2\theta} (E_x^2+E_y^2)] 
= \mp 4\beta \sin{2\theta} (E_x^2+E_y^2){n(e\tau)^3\over m\hbar^2[1+(\omega \tau)^2]}.
\end{equation}

Then, from Eqs.~\eqref{J_2} and~\eqref{f_1_infty} we obtain
\begin{multline}
- {e^2\tau\over \hbar}\qty(\bm E^\pm\cdot \bm \nabla_{\bm q})v_y^\pm 
 \qty[ \tau_\omega \sum_{\bm q} v_x^\pm  f_\pm^{(1)} (\bm q,\infty) +c.c.]
 = {e^3\tau\over \hbar^2}\qty(\bm E^\pm\cdot \bm \nabla_{\bm q})v_y^\pm  \sum_{\bm q} v_x^\pm 
 \qty(\bm E^\pm\cdot \bm \nabla_{\bm q}) f_0(\varepsilon_q)
\qty( \tau_\omega^2 +c.c.)
\\ 
 = -{e^3\tau\over \hbar^2}\qty(\bm E^\pm\cdot \bm \nabla_{\bm q})v_y^\pm  {n\over 2} \qty(\bm E^\pm\cdot \bm \nabla_{\bm q}) v_x^\pm 
\qty( \tau_\omega^2 +c.c.)
 = \mp 2\beta \sin{2\theta} (E_x^2+E_y^2){n(e\tau)^3[1-(\omega \tau)^2]\over m\hbar^2[1+(\omega \tau)^2]^2}.
\end{multline}

Let us now calculate the sums in Eq.~\eqref{J_2} containing $f_\pm^{(1,2)} (\bm q,0) $. They have a multiplier $v_x^\pm$ given by Eq.~\eqref{vx},
where the first term cancels at specular reflection as odd in $q_x$. Therefore we take all other terms at $\beta=0$, $\bm q = \bm k$:
\begin{align}
\qty(\bm E\cdot \bm \nabla_{\bm k})v_y^\pm &\sum_{\bm k} v_x^\pm f_\pm^{(1)} (\bm k,0) = \pm {2\beta\over m}\sin{2\theta}  E_y\sum_{\bm k} k_y f_\pm^{(1)} (\bm k,0),
\\
&\sum_{\bm k}v_x^\pm v_y^\pm f_\pm^{(2)} (\bm k,0) 
= \pm {2\beta\over m}\sin{2\theta}\sum_{\bm k}k_y^2f_\pm^{(2)} (\bm k,0) .
\end{align}
Then multiplying Eq.~\eqref{eq_f_1} at $\beta =0$ by $k_y$ and summing over $\bm k$ we obtain:
\begin{equation}
{1\over \tau_\omega}\sum_{\bm k} k_y f_\pm^{(1)} (\bm k,x) = -{e\over \hbar} \sum_{\bm k}k_y\qty(\bm E\cdot \bm \nabla_{\bm k})f_0(\varepsilon)
={e\over \hbar}E_y{n\over 2}.
\end{equation}
Similarly, multiplication of equation for $f_\pm^{(2)}$ (the steady-state version of Eq.~\eqref{kin_eq_E_field1}) by $k_y^2$ and summation over $\bm k$ at $\beta=0$ gives
\begin{equation}
{1\over \tau}\sum_{\bm k}k_y^2f_\pm^{(2)} (\bm k,x) = -{e\over \hbar}\sum_{\bm k}k_y^2\qty(\bm E\cdot \bm \nabla_{\bm k})f_\pm^{(1)} (\bm k,x) + c.c.
= 2E_y{e\over \hbar}\sum_{\bm k}k_y f_\pm^{(1)} (\bm k,x) + c.c.
= 2E_y^2\qty({e\over \hbar})^2 {n \tau \over [1+(\omega \tau)^2]}.
\end{equation}
Note that both $f_\pm^{(1)} (\bm k,0)$ and $f_\pm^{(2)} (\bm k,0)$ enter with the multiplier $E_y^2$. This means that they are the corrections arising in a response to the the field $\bm E \parallel y$, which is not sqreened by the electrons and, hence, coincides with that in the incident radiation~\cite{Durnev2020}.
The factor $E_x^2$ appears at  derivation of the corrections  $f_\pm^{(1,2)} (\bm k,\infty)$ far from the edge, where the electric field is also unscreened. Therefore both $E_x$ and $E_y$ driving the edge photocurrent coincide with those in the incident radiation.

Finally we obtain from Eq.~\eqref{J_2} the edge current in each spin subband
\begin{align}
J_{\rm edge}^\pm  = &\mp 4\beta \sin{2\theta} (E_x^2+E_y^2){n(e\tau)^3\over m\hbar^2[1+(\omega \tau)^2]}
 \pm 4\beta\sin{2\theta}  E_y^2  {n (e\tau)^3 \over m\hbar^2 [1+(\omega \tau)^2]} \nonumber
\\ &\mp 2\beta \sin{2\theta} (E_x^2+E_y^2){n(e\tau)^3[1-(\omega \tau)^2]\over m\hbar^2[1+(\omega \tau)^2]^2}
\pm 2\beta  \sin{2\theta}  E_y^2 {n(e\tau)^3[1-(\omega \tau)^2]\over m\hbar^2[1+(\omega \tau)^2]^2} \nonumber
\\ = &\mp 2\beta  \sin{2\theta}{n(e\tau)^3 \over m\hbar^2[1+(\omega \tau)^2]}
\qty[2E_x^2  + E_x^2{1-(\omega \tau)^2\over 1+(\omega \tau)^2} ]
=\mp 2\beta  \sin{2\theta}{n(e\tau)^3 [3+(\omega \tau)^2]\over m\hbar^2[1+(\omega \tau)^2]^2} E_x^2
.
\end{align}

\end{document}